\documentclass[twocolumn]{article}
\usepackage[T1]{fontenc}

\usepackage[english]{babel}
\usepackage{csquotes}

\usepackage[a4paper,top=2cm,bottom=2cm,left=2cm,right=2cm]{geometry}

\usepackage[authoryear, round]{natbib}

\usepackage{authblk} 

\usepackage{amsmath}
\usepackage{graphicx}
\usepackage[colorlinks=true, allcolors=blue]{hyperref}
\usepackage{libertine}       
\usepackage[libertine]{newtxmath} 

\title{PyTopo3D: A Python Framework for 3D SIMP-based Topology Optimization}
\author[1,3]{Jihoon Kim}
\author[2,3]{Namwoo Kang}

\affil[1]{Department of Mechanical Engineering, Korea Advanced Institute of Science and Technology, 291, Daehak-ro, Yuseong-gu, Daejeon 34141, Republic of Korea}
\affil[2]{Cho Chun Shik Graduate School of Mobility, Korea Advanced Institute of Science and Technology, 193, Munji-ro, Yuseong-gu, Daejeon 34051, Republic of Korea}
\affil[3]{Narnia Labs, 193, Munji-ro, Yuseong-gu, Daejeon 34051, Republic of Korea}
            
\date{}

\begin{document}

\maketitle

\begin{abstract}
Three-dimensional topology optimization (TO) is a powerful technique in engineering design, but readily usable, open-source implementations remain limited within the popular Python scientific environment. This paper introduces \texttt{PyTopo3D}, a software framework developed to address this gap. \texttt{PyTopo3D} provides a feature-rich tool for 3D TO by implementing the well-established Solid Isotropic Material with Penalization (SIMP) method and an Optimality Criteria (OC) update scheme, adapted and significantly enhanced from the efficient MATLAB code by \citet{liu2014TO}. While building on proven methodology, \texttt{PyTopo3D}'s primary contribution is its integration and extension within Python, leveraging sparse matrix operations, optional parallel solvers, and accelerated KD-Tree sensitivity filtering for performance. Crucially, it incorporates functionalities vital for practical engineering workflows, including the direct import of complex design domains and non-design obstacles via STL files, integrated 3D visualization of the optimization process, and direct STL export of optimized geometries for manufacturing or further analysis. \texttt{PyTopo3D} is presented as an accessible, performance-aware tool and citable reference designed to empower engineers, students, and researchers to more easily utilize 3D TO within their existing Python-based workflows.
\end{abstract}


\section{Introduction}

\begin{figure*}[tbh] 
  \centering
  \includegraphics[width=\textwidth]{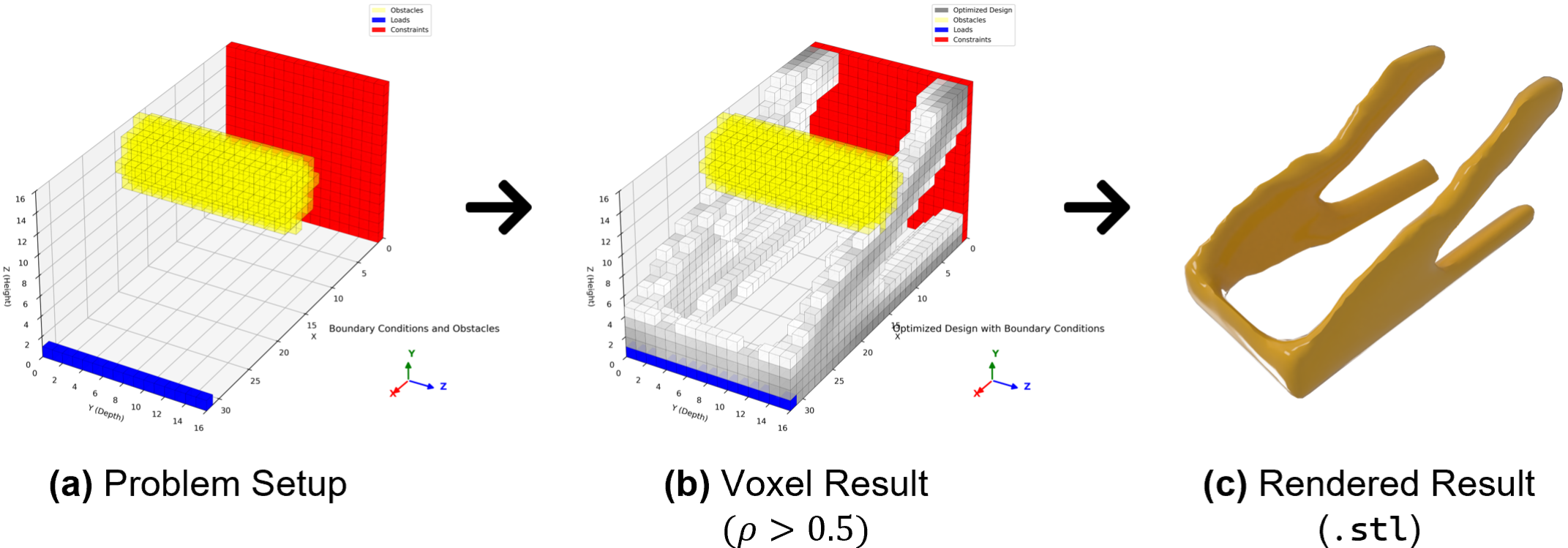}
  \caption{Optimization process and result for the 3D cantilever beam example using \texttt{PyTopo3D}: (a) Problem definition illustrating the design domain, boundary conditions, applied load, and an obstacle; (b) Optimized topology represented by voxels with pseudo-density above 0.5; (c) Rendered visualization of the final optimized structure.}
  \label{fig:cantilever_process} 
\end{figure*}


Topology Optimization (TO) has emerged as a powerful computational design methodology for determining the optimal distribution of material within a defined design space, subject to specified loads, boundary conditions, and performance objectives \citep{bendsoe2013topology, sigmund2013topology}. By enabling the generation of highly efficient, often non-intuitive structural layouts, TO finds widespread application in various engineering fields, including aerospace, automotive, and mechanical design. Its impact has been further amplified by the rise of additive manufacturing techniques capable of fabricating the complex geometries that TO often produces.

Among various TO approaches, the Solid Isotropic Material with Penalization (SIMP) method is particularly popular due to its conceptual simplicity, ease of implementation, and general effectiveness \citep{bendsoe1989optimal}. While numerous academic and educational implementations exist, particularly for two-dimensional problems and often within platforms like MATLAB \citep{sigmund200199, andreassen2011efficient}, there remains a persistent need for accessible, open-source, and computationally efficient tools specifically designed for three-dimensional (3D) TO within the burgeoning Python scientific ecosystem. Performing TO in 3D introduces significant computational challenges related to memory and processing time, and integrating TO capabilities smoothly into increasingly prevalent Python-based research and engineering workflows can present practical hurdles.

To address this gap, this paper presents \texttt{PyTopo3D}\footnote{Code available at: \url{https://github.com/jihoonkim888/PyTopo3D}}, a dedicated Python implementation for 3D topology optimization, adapted from the efficient MATLAB code presented by \citet{liu2014TO}. This code utilizes the well-established SIMP method coupled with the computationally efficient Optimality Criteria (OC) update scheme for compliance minimization under a volume constraint. \texttt{PyTopo3D} is designed with a focus on performance within the Python environment, leveraging sparse matrix operations, efficient finite element assembly techniques, optional parallel solvers, and optimized spatial filtering algorithms. Recognizing the needs of practical application, it incorporates features such as the ability to define complex design domains directly from Stereolithography (STL) files, include non-design (obstacle) regions, and export optimized results directly into STL format suitable for 3D printing or further CAD processing.

This paper serves primarily to document the \texttt{PyTopo3D} code, outlining its underlying methodology, core implementation details, key features, and providing guidance on basic usage. The principal aim is to furnish a stable, citable reference for researchers, students, and practitioners who utilize this software in their work, thereby facilitating reproducibility and proper attribution. The subsequent sections detail the software architecture and dependencies, the finite element formulation, the optimization algorithm and sensitivity filtering implementation, provide a basic usage example, discuss the code's perceived strengths and current limitations, and specify how to access and cite the software package.


\section{Methodology}

The topology optimization framework presented herein, \texttt{PyTopo3D}, is implemented entirely in Python and designed specifically for solving three-dimensional problems using the Solid Isotropic Material with Penalization (SIMP) method \citep{bendsoe2013topology}. The implementation prioritizes computational efficiency, modularity, and accessibility.

\subsection{Software Architecture and Dependencies}
\texttt{PyTopo3D} employs a modular architecture to enhance maintainability and extensibility. Key components include modules dedicated to the core optimization logic (\texttt{optimizer.py}), utility functions (\texttt{utils.py} for matrix assembly, filtering, sensitivity updates, linear system solving), input processing (\texttt{preprocessing.py}), result visualization (\texttt{visualization.py}), command-line interaction (\texttt{cli.py}), and experiment management (\texttt{runners.py}). The codebase utilizes a class-based structure promoting separation of concerns, with the primary optimization process orchestrated by the \texttt{top3d} function within the core module.

The software is implemented in Python (version 3.10 or later recommended) and builds upon core scientific libraries such as NumPy and SciPy for numerical computation and sparse linear algebra. Key functionalities including mesh processing and voxelization leverage libraries like Trimesh and scikit-image, while visualization capabilities are provided by Matplotlib. For accelerated computation, the framework interfaces with the optional PyPardiso parallel sparse solver, defaulting to SciPy's solvers otherwise to ensure broad compatibility. This design choice allows users to benefit from high-performance solving where available, without making it a strict requirement.

\subsection{Finite Element Analysis}
The structural analysis is performed using the Finite Element Method (FEM) based on an 8-node hexahedral (H8) element formulation. To maximize computational efficiency, particularly for large-scale 3D problems, several implementation strategies are employed. Element stiffness matrices are pre-computed, and global stiffness matrix assembly is accelerated using pre-calculated mapping indices (\texttt{edofMat}, \texttt{iK}, \texttt{jK}), enabling rapid construction within the optimization loop. The global stiffness matrix is stored and manipulated using SciPy's sparse matrix formats (COO initially, converted to CSR for solving) to minimize memory footprint and computational cost. The linear system of equations, \textit{\textbf{KU}} = \textit{\textbf{F}}, is solved using either the parallel direct solver from PyPardiso when available or fallback solvers within SciPy. Fixed displacement boundary conditions are applied directly by modifying the system matrix and load vector based on node indices, avoiding potential numerical issues associated with penalty methods.

\subsection{Optimization Algorithm: Optimality Criteria}
The material distribution is optimized using the well-established Optimality Criteria (OC) method, tailored for compliance minimization under a volume constraint. The SIMP approach is employed to relate the element-wise design variables (pseudo-densities, \(\rho_e\)) to the material's Young's Modulus (\(E_e\)), typically following \(E_e(\rho_e) = E_{min} + \rho_e^p (E_0 - E_{min})\), where \(E_0\) is the modulus of the solid material, \(E_{min}\) is a small non-zero modulus to prevent singularity, and \(p\) is the penalization power (commonly \(p=3\)).

The OC update scheme involves computing the element sensitivities of the objective function (compliance) with respect to the design variables. To ensure mesh-independence and prevent checkerboard patterns, these raw sensitivities are filtered using a spatial sensitivity filter, detailed below. The filtered sensitivities are then used within the OC update rule. The Lagrange multiplier corresponding to the volume constraint is efficiently found using a bisection algorithm. Standard move limits are applied during the design variable update to stabilize convergence. The volume constraint formulation correctly accounts for passive (non-design) regions, such as predefined obstacles.

\subsection{Sensitivity Filtering}
A spatial filter is applied to the element sensitivities to regularize the optimization problem. The filter weights are computed based on the distance between element centroids, decaying linearly with distance up to a specified filter radius (\(r_{min}\)). The filter operation can be expressed as a matrix-vector product, \(\tilde{\frac{\partial c}{\partial \rho_e}} = \textbf{H} \frac{\partial c}{\partial \rho_e}\), where \textbf{H} is the filter matrix. To construct \textbf{H} efficiently, neighbour searching is accelerated using SciPy's \texttt{cKDTree} implementation, which provides approximately O(log N) search performance for N elements. The filter matrix \textbf{H} is stored in a sparse format, allowing the filtering operation to be performed rapidly via sparse matrix multiplication.

\subsection{Workflow and Key Features}
A typical optimization process in \texttt{PyTopo3D} follows these steps: initialization of the design domain (potentially from an STL file), definition of loads and boundary conditions, and setup of optimization parameters; pre-processing, including computation of element stiffness matrices and filter matrices; iterative optimization loop involving FEM analysis, sensitivity computation, sensitivity filtering, design variable update using OC, and convergence checks; and finally, post-processing, including visualization and export of the optimized geometry.

Beyond the core 3D SIMP-OC implementation, \texttt{PyTopo3D} offers several features enhancing its utility. Performance is a key focus, achieved through NumPy vectorization, efficient sparse matrix handling, optional parallel solving via PyPardiso, and accelerated filter construction. The code supports the inclusion of fixed non-design (obstacle) regions within the design domain. Furthermore, complex design domains can be initialized directly from STL geometry files, enabling optimization within non-rectangular boundaries. Integrated visualization tools allow for monitoring the optimization progress in 3D and generating animations. Final designs can be exported as STL files suitable for additive manufacturing or integration into CAD workflows. The framework also includes utilities for managing and tracking multiple optimization runs, facilitating parameter studies.

\section{Basic Usage Example}

To illustrate the functionality of \texttt{PyTopo3D}, a standard 3D cantilever beam problem is considered. The design domain is discretized into 32 x 16 x 16 hexahedral elements. A fixed boundary condition is applied to the left face, and a vertical point load is applied at the center of the free end's bottom edge. A cylindrical obstacle region is defined within the domain where material is disallowed. The optimization aims to minimize compliance subject to a volume fraction constraint of 20\%.

The optimization is executed using \texttt{PyTopo3D} with the aforementioned parameters and the Optimality Criteria solver. The process converges after N iterations. The complete process, from problem definition through optimization to the final result visualization, is summarized in Figure~\ref{fig:cantilever_process}. Specifically, Figure~\ref{fig:cantilever_process}(a) illustrates the problem setup including the design space, boundary conditions, load application point, and the obstacle geometry. Figure~\ref{fig:cantilever_process}(b) shows the resulting optimized topology represented by voxels with pseudo-density above 0.5, clearly forming structural members that distribute the load back to the fixed support while avoiding the obstacle region. Finally, Figure~\ref{fig:cantilever_process}(c) provides a rendered visualization of the optimized structure, suitable for additive manufacturing or further analysis.


\section{Discussion}

The \texttt{PyTopo3D} implementation presented offers several notable strengths, positioning it as a valuable tool within the topology optimization landscape, while also having limitations inherent to its current scope and design choices. A primary advantage lies in its implementation within the Python programming language. This leverages the extensive and rapidly growing scientific computing ecosystem available in Python, facilitating integration with other simulation, data analysis, or machine learning workflows commonly employed in research and engineering. Consequently, \texttt{PyTopo3D} provides an accessible alternative for users, including students and researchers, who may be less familiar with traditional platforms like MATLAB, potentially broadening the adoption and exploration of topology optimization techniques. Despite being an interpreted language implementation, careful attention to performance optimization --- including NumPy vectorization, efficient sparse matrix handling, and optional parallel solvers --- allows the code to remain computationally competitive for many moderately sized three-dimensional problems encountered in practice. Furthermore, practical features such as support for non-design obstacle regions, the ability to define complex design domains directly from STL files, and the direct export of optimized results to STL format enhance its applicability beyond simple academic benchmarks towards more realistic engineering design tasks.

\subsection{Enhancements Over Foundational Code}

While \texttt{PyTopo3D} adapts the core SIMP/OC methodology from \citet{liu2014TO}, it incorporates substantial architectural, functional, and usability enhancements, transitioning it towards a more feature-rich engineering tool suitable for practical applications.

\subsubsection{Architecture and Performance}
Architecturally, the implementation was refactored from a monolithic script into a modular Python package, separating concerns related to optimization, utilities, preprocessing, visualization, and experiment management. This modularity, combined with full integration into the Python scientific stack (NumPy, SciPy, Matplotlib), enhances maintainability and extensibility. Performance optimizations represent a key focus; sensitivity filter construction was significantly accelerated using KD-Trees for neighbor searching, replacing potentially slower methods. Furthermore, the framework leverages optional parallel linear system solving through PyPardiso, utilizes pre-computed element mapping indices for faster stiffness matrix assembly, and employs vectorized NumPy operations extensively.

\subsubsection{Expanded Functionality}
Functionality has been considerably expanded beyond the original scope to address practical engineering scenarios. \texttt{PyTopo3D} introduces critical support for defining non-design obstacle regions within the optimization domain and allows the import of complex design domains directly from STL files, moving beyond simple rectangular domains. Post-processing is streamlined via direct export of the optimized topology to STL format, suitable for additive manufacturing or CAD import. Visualization capabilities are also enhanced with interactive 3D views, optional animation of the optimization process, and graphical representation of boundary conditions and loads.

\subsubsection{Usability and Workflow Integration}
Usability and integration within modern workflows are improved through a comprehensive Command-Line Interface (CLI) and a Python API, enabling use both as a standalone application and as a library within larger Python projects. Configuration is made more flexible, for instance, through JSON files for complex obstacle definitions. The user experience is further enhanced by a detailed logging system, real-time progress reporting during optimization, and integrated experiment management utilities for tracking runs and results. Enhanced inline documentation and help text also contribute to usability.

\subsubsection{Technical Refinements and Additional Features}
Several technical refinements contribute to robustness and efficiency. These include a more efficient sparse matrix-based filter implementation, improved memory management particularly beneficial for large problems, enhanced tracking of the objective function convergence, and more flexible setup options for boundary conditions. The adoption of Python type annotations improves code clarity and aids long-term maintenance. Additional features further increase practical utility, such as tools for results analysis, optional mesh smoothing capabilities for exported STL files, improved handling of volume constraints in the presence of obstacles, and the capability to generate animations of the optimization process.

Collectively, these developments represent a significant evolution from the original MATLAB code, broadening the applicability and performance of the topology optimization approach within the Python ecosystem.

\subsection{Performance Comparison}

To quantitatively assess the computational performance of \texttt{PyTopo3D}, a benchmark was conducted comparing it against the foundational MATLAB code \citep{liu2014TO} using an identical problem setup and hardware environment. Both implementations solved an 8192 element (32x16x16) cantilever beam problem for 200 iterations on the same laptop running Windows 11 and equipped with an Intel Core Ultra 7 155U processor. Standard optimization parameters were used for this benchmark, specifically a volume fraction (\texttt{volfrac}) of 0.2, a SIMP penalization power (\texttt{penal}) of 3.0, and a filter radius (\texttt{rmin}) of 4.0 elements.

The \texttt{PyTopo3D} implementation demonstrated significantly faster execution, completing the optimization in approximately 147.3 seconds, compared to 274.3 seconds for the MATLAB version, achieving a speedup of roughly 1.86 times for this specific test case.

An analysis of the time distribution among key computational phases, presented in Table~\ref{tab:phase_timing_condensed}, highlights the sources of this performance difference. The most substantial time saving in \texttt{PyTopo3D} comes from the linear system solve phase, which required only 99.8 seconds compared to 234.1 seconds in MATLAB. This efficiency likely stems from the use of optimized sparse solvers available in SciPy or optional parallel solvers such as as PyPardiso. In contrast, the matrix assembly phase took longer in the Python implementation (33.9 seconds) than in MATLAB (24.0 seconds). The time spent on the sensitivity filtering and Optimality Criteria update steps was broadly comparable or slightly faster in absolute terms for \texttt{PyTopo3D}.

\begin{table}[tbh]
\centering
\caption{Phase Timing Comparison (Seconds and \% of Total Time)}
\label{tab:phase_timing_condensed}
\setlength{\tabcolsep}{4pt} 
\begin{tabular}{lrrrr}
\hline
          & \multicolumn{2}{c}{MATLAB} & \multicolumn{2}{c}{\texttt{PyTopo3D}} \\ \cline{2-5}
Phase     & Time (s)      & (\%)     & Time (s)        & (\%)        \\ \hline
Assembly  & 24.0         & 8.7             & 33.9      & 23.0          \\
Solve     & 234.1        & 85.4            & 99.8   & 67.8       \\
Filter    & 0.8          & 0.3             & 0.7       & 0.5           \\
Update    & 13.9         & 5.1             & 10.8      & 7.4           \\ \hline 
\textbf{Total} & \textbf{274.3} & \textbf{100.0} & \textbf{147.3} & \textbf{100.0} \\ \hline 
\end{tabular}
\end{table}

In summary, the benchmark results executed on the same hardware confirm that the \texttt{PyTopo3D} framework offers a substantial computational speed advantage over the foundational MATLAB code for this 3D topology optimization problem with typical parameters, primarily driven by significant efficiencies gained in the linear system solve step within the Python environment leveraging modern numerical libraries.

\subsection{Limitations and Future Work}

Future development efforts for PyTopo3D could concentrate on expanding its capabilities beyond the current focus on linear elastic compliance minimization to encompass a broader range of physics, potentially including thermal analysis or coupled multi physics problems. Additionally, addressing the scalability challenges inherent in very large scale 3D computations remains an important direction, possibly through the exploration of alternative numerical strategies or further computational optimizations to reduce memory consumption and processing time.


\section{Conclusion}

This paper presented PyTopo3D, an open-source Python framework providing an accessible, efficient, and crucially, feature-rich tool for 3D SIMP-based topology optimization. By substantially enhancing the foundational MATLAB code from \citet{liu2014TO}, PyTopo3D delivers a performance-optimized implementation tailored specifically for the Python scientific ecosystem. Its key contribution lies in integrating vital practical functionalities essential for engineering workflows, such as direct handling of complex geometries and obstacles via STL files, alongside usability features like a dual CLI/API and experiment management. PyTopo3D significantly lowers the barrier for engineers, students, and researchers aiming to leverage powerful 3D topology optimization techniques directly within their Python environments. While current work focuses on linear elastic problems, future efforts may broaden the physics scope and further optimize performance, building upon this documented, accessible foundation designed to foster wider adoption in the computational design community.


\section*{Code Availability} 

The \texttt{PyTopo3D} source code described in this paper is publicly available.
The primary development repository is hosted on GitHub at:
\url{https://github.com/jihoonkim888/PyTopo3D}.


\bibliographystyle{apalike}
\bibliography{references}

\end{document}